SOFTWARE TOOL ARTICLE

# The AMPEL Platform: An Open-Source Project for Developing Hospital-Wide Clinical Decision Support Systems under EU-Regulation

Alexander Twrdik[1†], Daniel Steinbach[1†*], Martin Federbusch[1], Clemens Malkawi[1], Maria Schmidt[1†], Toralf Kirsten[2†]

1 University of Leipzig Medical Center, Department for Clinical AI and Translational Medicine, Leipzig, Germany

2 University of Leipzig, Department Medical Data Science, Leipzig, Germany

† These authors contributed equally to this work.

* ampel@medizin.uni-leipzig.de

## Abstract

As incentives to digitize more and more aspects of medical healthcare are gaining traction, additional medical data is becoming available. A major goal is to use this data to gain further insights, support medical staff, and ultimately improve patient care. This demands sophisticated solutions that can handle current and future vast amounts of data, all without disrupting well-established workflows of care takers. Additionally, the solution must fulfill the documentation and risk-management obligations of the Medical Device Regulation (MDR), which governs whether a device may be used in patient care in Europe.

In this context, we introduce the AMPEL platform, an open-source solution for creating clinical decision support systems (CDSS) that fulfil the requirements of the MDR. Centered around a message broker (Kafka), it adopts an event-driven, distributed architecture, allowing for scaling and efficient processing of medical data as it gets generated. Relying on platform-specific events to implicitly describe the platform's interfaces makes it both extensible and agnostic to the hospital information system in use. With these events being defined in the Protocol Buffers format, medically evaluated algorithms can be independently implemented in a variety of programming languages and easily linked to the platform, extending its analytic capabilities as desired. The platform features a basic UI and alert functionality that can be used to gain medical staff's attention and present analysis results in a traceable and explainable fashion.

We further show in a use case how the platform was used to build LAMPE, a CDSS developed and operated at University of Leipzig Medical Center. It demonstrates how the platform solution can be used in a complex and already established environment, while acknowledging current limitations and outlining planned developments to overcome these.

## Introduction

Delays in the detection of diseases and complications are, unfortunately, part of everyday clinical care. Accordingly, numerous initiatives seek to improve patient safety by reducing missed complications and to advance diagnostic innovation by detecting the previously undetectable. Clinical decision support systems (CDSS) have a key role in this context by assisting healthcare professionals through automated data processing.[1]

In this regard, CDSS use a number of different approaches. The software described here is intended to autonomously monitor the entire patient record, encompassing all structured patient data available in the clinical systems, and to generate proactive alerts. This monitoring software can be thought of as an additional safety net spanning the entire course of clinical care.

We identified the following characteristics as integral and beneficial to such a system. Clinicians must not enter any additional data. Calculations and notifications are event-driven: they occur as soon as the underlying data change, without requiring any further triggers, such as opening the patient's chart or launching the software, as is common with commercially available products[2] or interoperability standards[3] (see Figure 1).

Real-time data processing is another relevant parameter. Time-critical conditions (e.g. those relevant to the early detection of sepsis) must trigger alerts as quickly as possible.[4] The hospital-wide platform must be designed for openness and scalability, enabling numerous disease-specific algorithms to operate concurrently and to be replaced on an ongoing basis. Enabling multicenter studies also requires the software to function regardless of which hospital information system (HIS) is used. A comprehensive list of our requirements is provided in the *stakeholder* and *system requirements* of the accompanying quality management system (QMS) documents (https://ampel-wiki.eu).

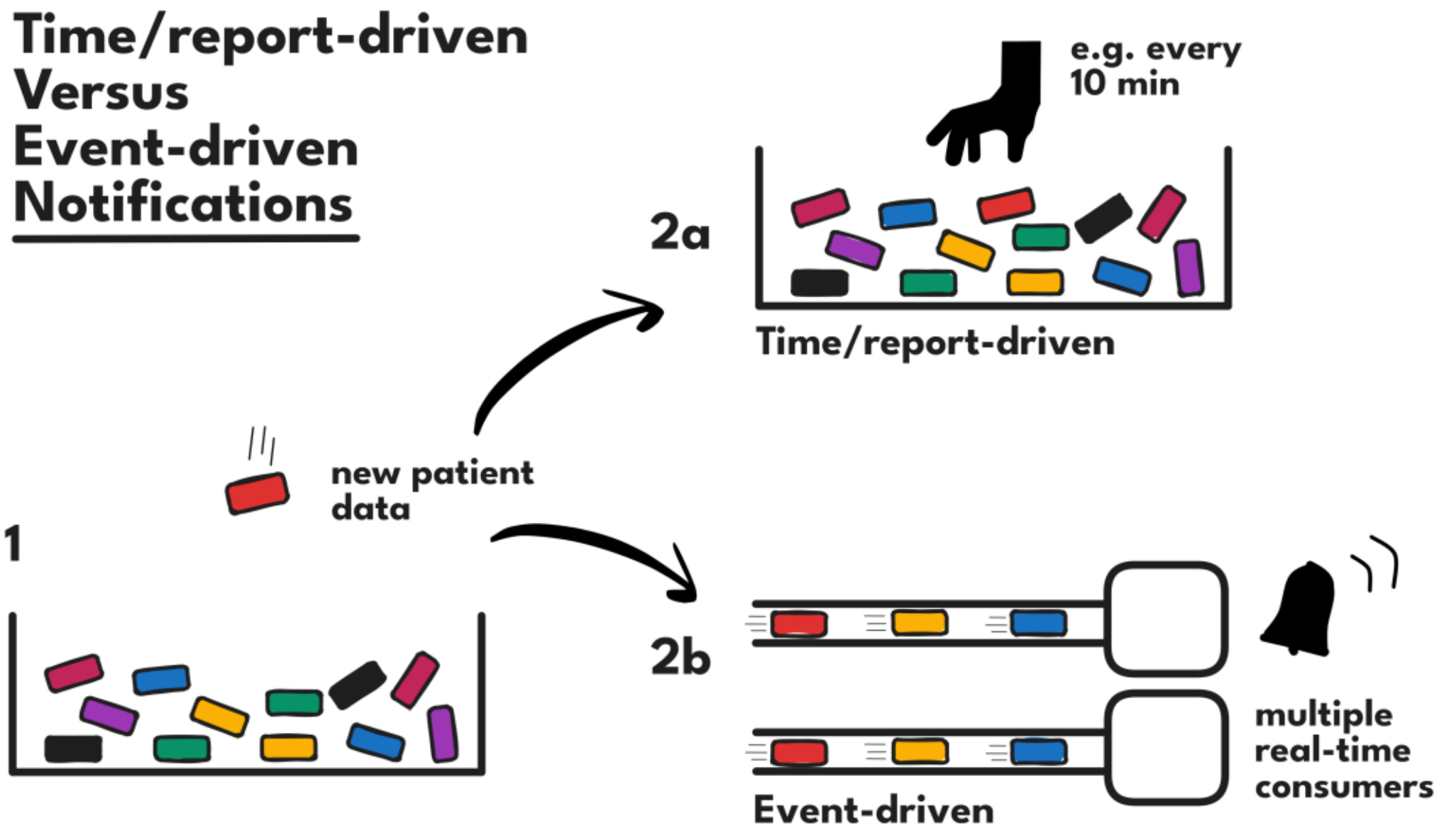

***Figure 1.*** *Time- or report-driven solutions (2a) run algorithms only at fixed intervals or in response to a user action (clicking a button or opening a patient's chart). By contrast, the event-driven software (2b) described here uses the real-time data themselves as triggers. The algorithms run in parallel as independent services, while individually processing the same events.*

Despite high expectations and numerous promises, there is a fundamental evidence gap regarding the effectiveness of AI in patient care. Although hundreds of new AI medical products are currently made available on the EU market each year, the proportion of products with evidence of effects on patients remains below 1%.[5,6] This evidence gap also stems from the fact that, beyond a certain level of translational progress, research projects in the clinical decision support field encounter not only a technical barrier but also a regulatory one. In Europe, effects on patients may only be studied by using medical device software (MDSW), which falls under the Medical Device Regulation (MDR). The regulatory requirements therein are associated with a substantial effort and, to the best of our knowledge, have not previously been met within an open-source software (OSS) project. The aim of the AMPEL platform is to overcome this barrier. The platform is intended to help other healthcare institutions develop in-house MDSW. The resulting CDSS can be fully used in patient care and thus also meets the requirements for clinical studies evaluating effects on patients.

This paper describes the technical AMPEL platform, which can serve as a shared foundation for such an hospital-specific CDSS. It also demonstrates the first successful implementation at the University of Leipzig Medical Center (ULMC) in Germany.

# Methods

## Implementation / Architecture

### Distributed System

The introduced platform is meant to be primarily utilised by facilities that are providing tertiary care. This mainly includes hospitals with several specialist departments and larger quantities of patients, which also results in large quantities of diverse data. This data needs to be permanently ingested into the platform at high volumes. Derived information must also be queried afterwards. In such a scenario it's foreseeable that a monolithic application would not scale well. Thus, the platform is designed as a distributed system such that its components are closed entities serving a single domain aspect (going forward these components are called services).

Services have to communicate via a message broker, effectively ensuring that message producers and consumers are operationally decoupled. This, in turn, minimizes backpressure within the platform while maximizing throughput, which enables near realtime data processing. All services communicating via the message broker are using a choreographed approach (as opposed to an orchestrated one). In this scenario all services have to know about available and required messages to serve specific use cases. Doing so introduces a higher complexity in terms of developing and operating services. At the same time, usage of a dedicated communication orchestrator is rendered unnecessary, removing a potential single-point-of-failure within the system. The platform actively forgoes usage of a data warehouse. These are mainly designed for read-heavy workloads whereas the platform requires solutions for write-heavy workloads in order to handle high volumes of data ingestion. Instead, the platform relies on services using dedicated Online Transaction Processing (OLTP) solutions. Doing so enables the use of a fitting database for each service's domain scope and use case at hand. Additionally, it keeps the data close to where it's needed, minimizing latency. A high-level overview of the described architecture is shown in figure 2.

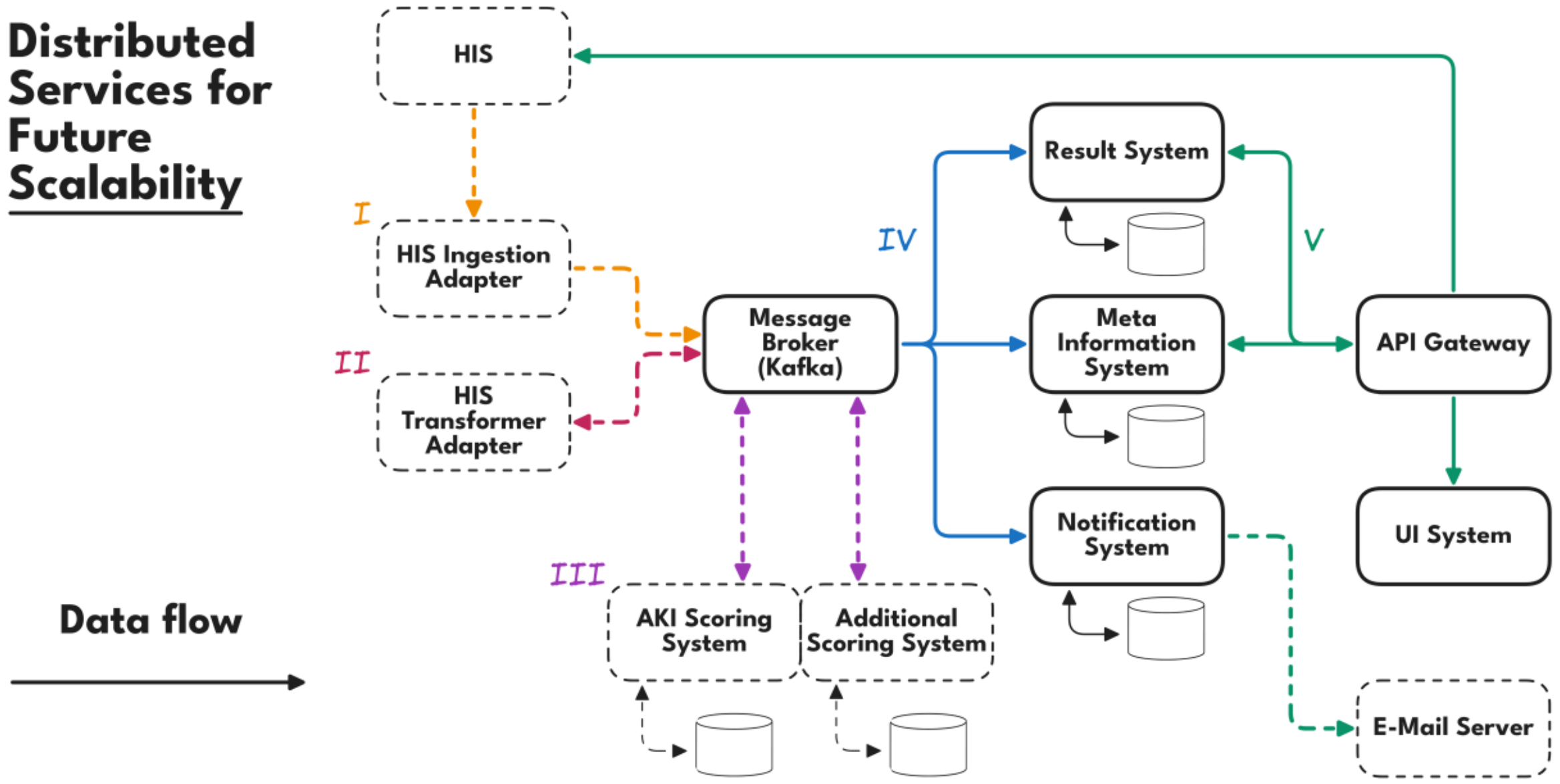

***Figure 2.*** *The platform's architecture is depicted as the overall dataflow, with core components displayed in solid lines. External components, necessary to turn it into a CDSS, are displayed in dotted lines.* ***I)*** *Ingesting raw HIS data;* ***II)*** *transforming data into platform events;* ***III)*** *calculating scores by reading and analyzing data;* ***IV)*** *reading platform events and using them in a domain specific fashion;* ***V)*** *making data available to external systems.*

### Kafka as the communication channel (Message Broker)

As already described, the platform requires a message broker in order to enable services to communicate with each other. Within the scope of the platform, a message broker needs to meet the following requirements: (i) guarantee message delivery to avoid losing data leading to inconsistencies in downstream services, (ii) keeping message order to rule out logical inconsistencies that may lead to wrong calculations in consumers and eventually, wrong decision support alerts, (iii) guarantee message durability such that messages are not lost even in case of a system crash, (iv) retaining messages to allow for replayability when consumers are dynamically added allowing for retrospective message consumption.

The AMPEL platform makes use of Apache Kafka to meet these requirements.[7,8] It offers an at-least-once message delivery guarantee, i.e. a message can be delivered multiple times, but will be delivered at least once. Dealing with an at-least-once delivery guarantee requires consumers to handle messages in an idempotent fashion. The platform's services ensure this by employing a transactional inbox. This part takes care of definitively processing, but also de-duplicating incoming messages. The latter trait effectively renders the whole process idempotent.

Kafka allows for the grouping of messages in a logical way using so-called topics. On a technical level these are made up of several partitions, on which message ordering is guaranteed only. Doing so is sufficient for the platform as it can map a single medical encounter to a single Kafka partition using a unique identifier as part of the message itself (medical encounter ID). Message ordering however, is irrelevant across multiple medical encounters since this is equivalent to processing independent entities in parallel. The ordering only needs to be strictly guaranteed within each medical encounter.

Usage of multiple topics is deliberately avoided. Otherwise, a message consumer would be responsible for ensuring message ordering since reading from multiple topics introduces out-of-order messages. This happens, amongst other factors, due to network latency. It would be possible to deal with these effects using advanced and sophisticated streaming techniques like windowing.[9,10] However, doing so requires introducing a complexity, which is not proportional to the benefits. Thus, the platform keeps all its messages in a single topic.

As a solution that can be operated in a clustered fashion, in combination with append-only logs and strict producer acknowledgement modes, Kafka ensures message durability. This actively minimizes data loss even in case of a system outage. In addition, it can retain messages based on a configured duration. This approach allows for consumers being dynamically added to the platform with the option of consuming messages retrospectively. Furthermore, it enables self-healing capabilities of the platform after partial system outages as consumers can eventually catch up to the latest messages.

### Protocol Buffers as the Message Format

Due to services communicating with each other in a choreographed manner, there are inherently no explicit interfaces in place. Instead, these are implicitly defined via the messages that services exchange. Thus, the platform requires a format for defining messages that: (i) can validate messages to avoid introducing incorrect messages (poison pills) to the system that have the potential to bring down downstream services, (ii) is programming language agnostic, so that the platform can support different ecosystems, (iii) is efficiently serializing and deserializing messages to keep latency to a minimum, and (iv) allows adjusting message structures while keeping producers and consumers compatible.

These requirements are met by making use of Protocol Buffers *(Protobuf)*. It allows for message structures being defined via a dedicated language (Protobuf Language). A compiler translates them into usable constructs for specific programming languages. Compared to common formats like JSON or XML, Protobuf is also more efficient regarding serialization and deserialization.[11] Additionally, Protobuf comes with support for schema evolution, a procedure allowing for changes in message structures while keeping both producers and consumers compatible. This trait is vastly beneficial for operating the platform. Having schema evolution support allows for iteratively adjusting producers and consumers keeping downtime to a minimum. Combining Protobuf with an external schema registry also allows for messages being validated. This effectively acts as an additional layer securing the platform from malformed messages (poison pills) that can corrupt downstream services.

Listing 1 shows the structure of the platform's root message. It is described in such a way that the structure forms an envelope. I.e. an actual message is going to be embedded in this overarching message, which carries general metadata. This information allows the platform to route messages without having to deserialize the whole message. Following this approach leads to minimizing latency within the platform.

Protocol Buffer

```
message AmpelMessage {
  string event_id = 1;
  google.protobuf.Timestamp timestamp = 2;
  string partitioning_key = 3;
  string emitter = 4;
  string event_type = 5;
  // keeping space for future metadata additions
  google.protobuf.Any payload = 10;
}
```

***Listing 1.*** *The platform's root message, defining an envelope for other messages to be embedded in.*

### Events as Messages

Next to the message format, we decided to go with events in terms of message content, integration events to be specific. An event itself simply describes something that has already happened and needs to be broadcasted to interested parties as a fact. This approach is a natural fit for an event-driven architecture and leads to a small message footprint, as they only carry information describing an event. The events themselves are curated in such a way that consuming parties never have to ask other parties for additional details, i.e. no reference links to resources are being used. This limits roundtrips within the platform and avoids flooding services with requests.

Numerous events are defined within the platform's current state using Protobuf (e.g. *AmpelMedicalCaseCreatedEvent*, *AmpelPatientDiedEvent*). For the exhaustive list see the technical manual in the accompanying online wiki (https://ampel-wiki.eu). These events can be used by other parties trying to integrate the platform into their existing environment. This is mainly useful for services bridging an existing HIS and this platform, as well as independent scoring systems being developed.

Within the medical domain's scope it is otherwise common to describe exchanged information using Fast Healthcare Interoperability Resources (FHIR).[12] However, this standard is not usable within the event-driven platform. The main reason being that FHIR resources are always described in their current state, which would require every consumer to keep a reference resource in order to calculate a difference for inferring semantic meaning. Doing so would introduce unnecessary complexity and computational overhead, degrading the platform's latency compared with the proposed solution.

### HIS agnostic

Defining the platform's interface via curated integration events, makes it entirely HIS agnostic. However, this requires a translation layer between the two systems. Such a layer is responsible for effectively taking information coming from the HIS and translating it into platform events. This approach is shown in figure 2 (subitem I). It is advised to keep this process split into two parts. The first one simply stores unaltered information coming from the HIS within a dedicated Kafka topic. Doing so has three advantages: (i) it retains raw information from the HIS, allowing for alternative translation implementations in the future, (ii) it prevents losing messages in case of a system outage, and (iii) the throughput is maximized since no transformative or analyzing steps are performed on the data. The second part reads this information back, translates it into platform events and eventually feeds these into the platform via the dedicated AMPEL Kafka topic. This split approach shields the translation part against information flooding, i.e.during peaks of high volume data influx coming from the HIS.

The translation layer is not part of the platform itself and needs to be added following the requirements of where the platform is intended to be integrated.

### Extensibility Through Independent Scoring Systems

All aspects introduced so far form the scaffold of the AMPEL platform. It does not, however, include any scoring system itself. These would need to be added independently, according to the intended use of the constructed CDSS. An example for such a scoring system is AKI, which detects an acute kidney injury by analyzing a combination of a patient's laboratory data as well as clinical procedures and patient metadata.[13] It does so, like any other potential scoring system, by processing relevant platform events as received via the message broker in order to calculate a score. Irrelevant events are effectively filtered out by making use of the introduced message envelope’s metainformation.

A score calculated by any scoring system is defined as the severity of a finding within analyzed data. Instead of using a numerical value for this purpose, we opted for ***red** (anomaly with high severity)*, ***yellow** (anomaly with moderate severity)* and ***green** (no anomaly found)* as discrete values. This comes with the advantage that any complex categorization / mapping is happening within the scoring system itself. Furthermore, it allows for downstream visualizations to be grasped at a glance, which is crucial in a care related environment.

The score calculated by a scoring system is subsequently reported back to the platform as a dedicated event via the message broker as well. Dedicated platform services, as depicted in figure 2 (subitem IV), will then utilize it in order to trigger CDSS-specific functionalities (if needed). These include notifying hospital staff via e-mail or providing score information for the HIS. Following this approach, scoring systems do not need to run periodically. Instead, they are being triggered by events. This minimizes load on the overall system as analyses are only carried out if there is a need for it.

We provide an SDK (Software Development Kit) to ease implementing such scoring systems and integrating them into the platform. It already covers common tasks like event handling within the platform and also provides an implementation of a transactional inbox. The SDK is currently available for Java, but other programming languages will be supported in the future.

### Notifications

Simply having scoring systems analyze data and produce scores does not in itself provide any added value. Scores need to reach hospital staff so that they can take further action based on the transmitted information. For this purpose, the platform provides a dedicated notification service that is responsible for sending e.g. e-mails. It reads events related to scores and patients from the platform's *AMPEL* topic within Kafka using a transactional inbox. The patient events are solely used for capturing context information that can be used in notifications. They are stored in a dedicated database for later easy retrieval.

Whenever a *red* or *yellow* score as part of an event is being observed, a process is triggered to send out an e-mail as a notification. This includes retrieving any previously stored patient information and using it to enrich the score. The resulting information is then used to populate a configurable template for both the e-mail subject and body. This allows for easily customizable notifications without requiring changes to the service itself.

With alerting humans being the crucial task of the platform, we needed to ensure that notifications were sent out at least once. This was accomplished by processing a score, including information enrichment and sending out the notification itself, as part of a single transaction. An approach that was also combined with a dedicated retry logic.

### Visualization

Aside from actively providing information to a user via notifications, the platform also offers a basic user interface. Since it doesn't come with a navigation itself, it is intended to be integrated into a HIS as a linked external view. This view is centered around a medical encounter and displays the latest scores of all scoring systems for this encounter as well as associated patient metainformation. In order to aid users in identifying the most important information at a glance, the interface groups entries using vibrant colors (figure 3).

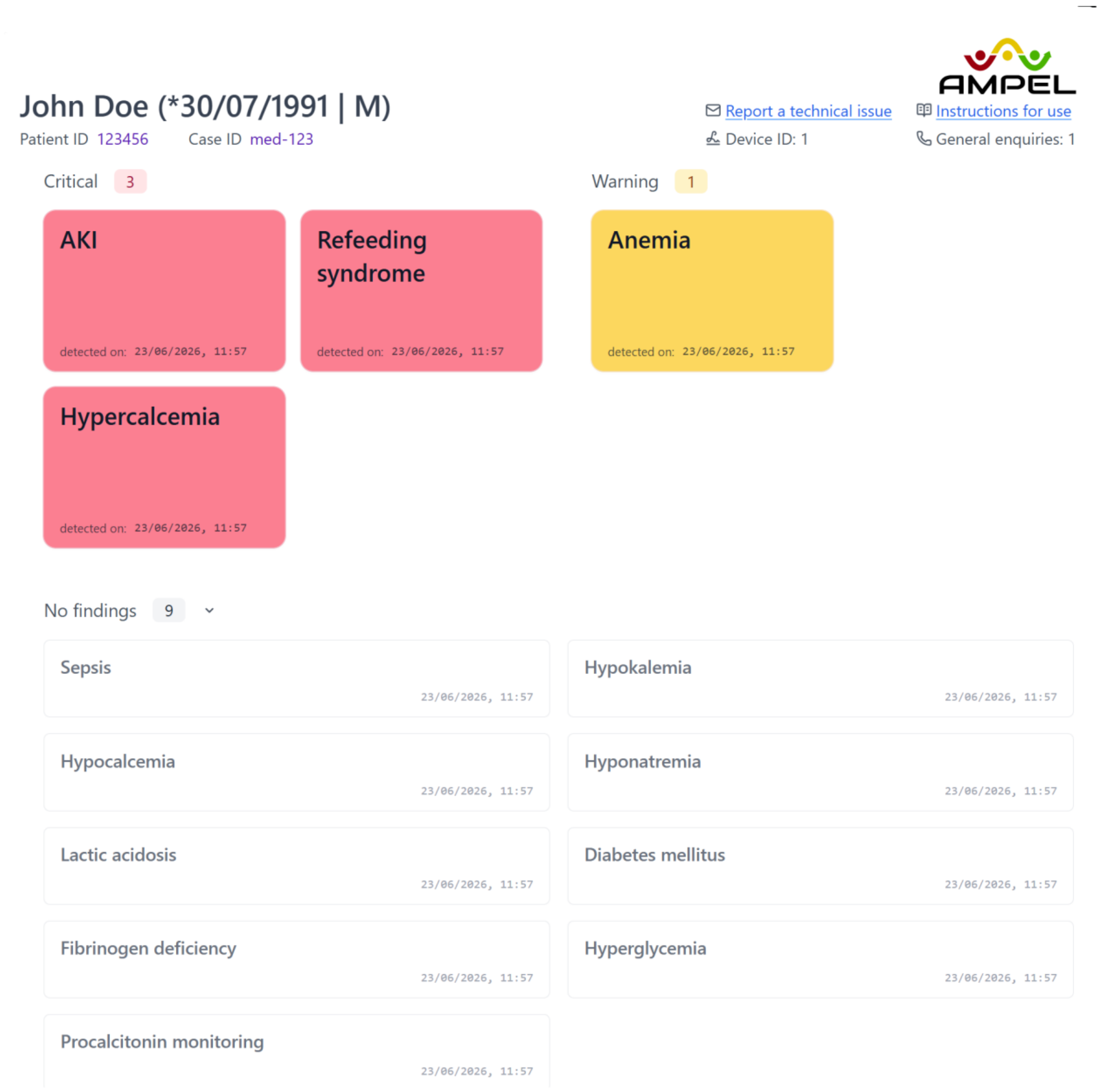

***Figure 3.*** *Basic user interface centered around an example medical encounter. It displays the latest scores from internal scoring systems. Scores are color-coded for a focussed assessment of severity. Calculated "green scores" (no anomaly found) are deliberately displayed without color to prevent unnecessary distractions.*

For further insights the interface also provides a detailed view, which appears when selecting any of the scoring systems. This detailed view is depicted in figure 4 and includes information on how the scores of a single scoring system were calculated, a history of results and decision support comments (recommended course of action).

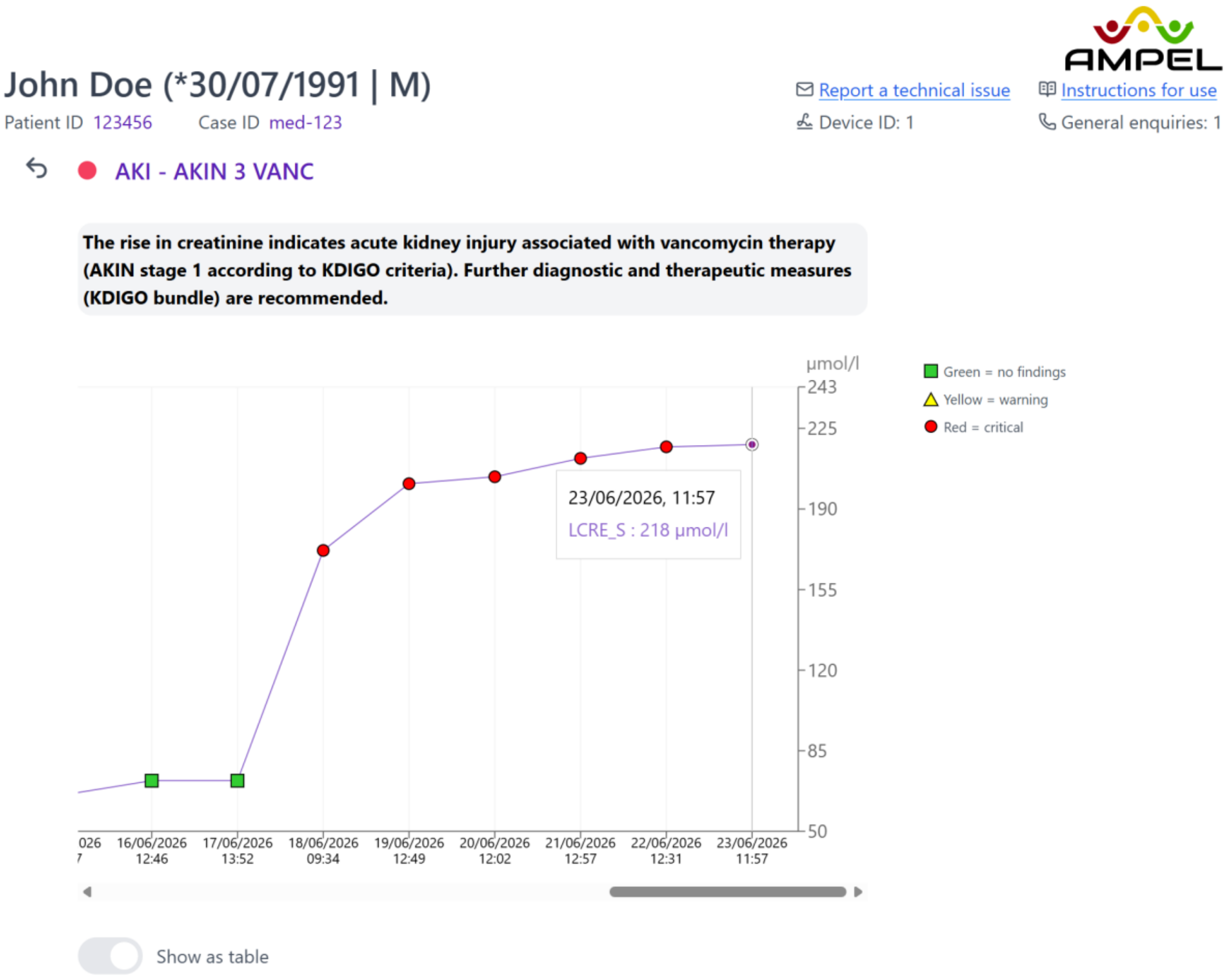

***Figure 4.*** *Detailed view with the history of results for the AKI scoring system in an example medical encounter. This also includes a decision support comment with a recommended course of action based on the latest result of this scoring system.*

## Operation

### Documentation

As already mentioned, the platform has to be actively turned into a CDSS. This is mainly done by providing one or more scoring systems as well as a layer for ingesting data coming from a HIS. For this purpose we provide a dedicated wiki (https://ampel-wiki.eu) containing a detailed manual on how to do this on a technical level. Furthermore, it also offers a regulatory manual for creating a medical device, as well as MDR-compliant quality management documents. The information found there is meant to aid in the development of medical devices at other locations and provides further details on the AMPEL platform.

### Hardware Requirements

In order to operate the platform reliably, an Apache Kafka cluster with at least 3 distinct nodes is required. This constellation is necessary to protect the message broker against potential data loss and keep the platform operable in case of partial outages. Services already included in the platform require, in total, 4 CPU cores and 8GB RAM. Furthermore, a container runtime needs to be available to run these platform services, as they are provided as Docker images.

By creating a CDSS based on the platform, implementing different scoring systems as well as creating the HIS translation layer need to be taken into account. Requirements for these activities cannot be provided here as they are highly dependent on the individual solution.

## Ethics statement

Ethical approval was not required for this study, as it did not involve human participants, animal subjects, or identifiable personal data.

## Consent statement

Consent is not applicable, as this manuscript does not contain any individual data in any form.

## Study registration

Study registration is not applicable.

# Use case

Based on the proposed platform, we created a CDSS as a medical device that is currently operated at ULMC, called ***LAMPE***. Throughout the operational timeframe of six months medical staff were alerted to 3755 detected anomalies of different severities with regards to an acute kidney injury. The clinical relevance of these anomalies will be examined as part of a separate publication. For context, the above solution had to be integrated into an environment running *SAP i.s.h.med* as the HIS. Visualizing results within SAP itself is not part of the created medical device but was customized through SAP operators on-site. Additionally, notifications happened through a shared e-mail inbox to a pharmaceutical intervention team, which the CDSS was configured to use as an alert channel. Figure 5 depicts the resulting architecture of LAMPE and the following sections describe in more detail how it was implemented.

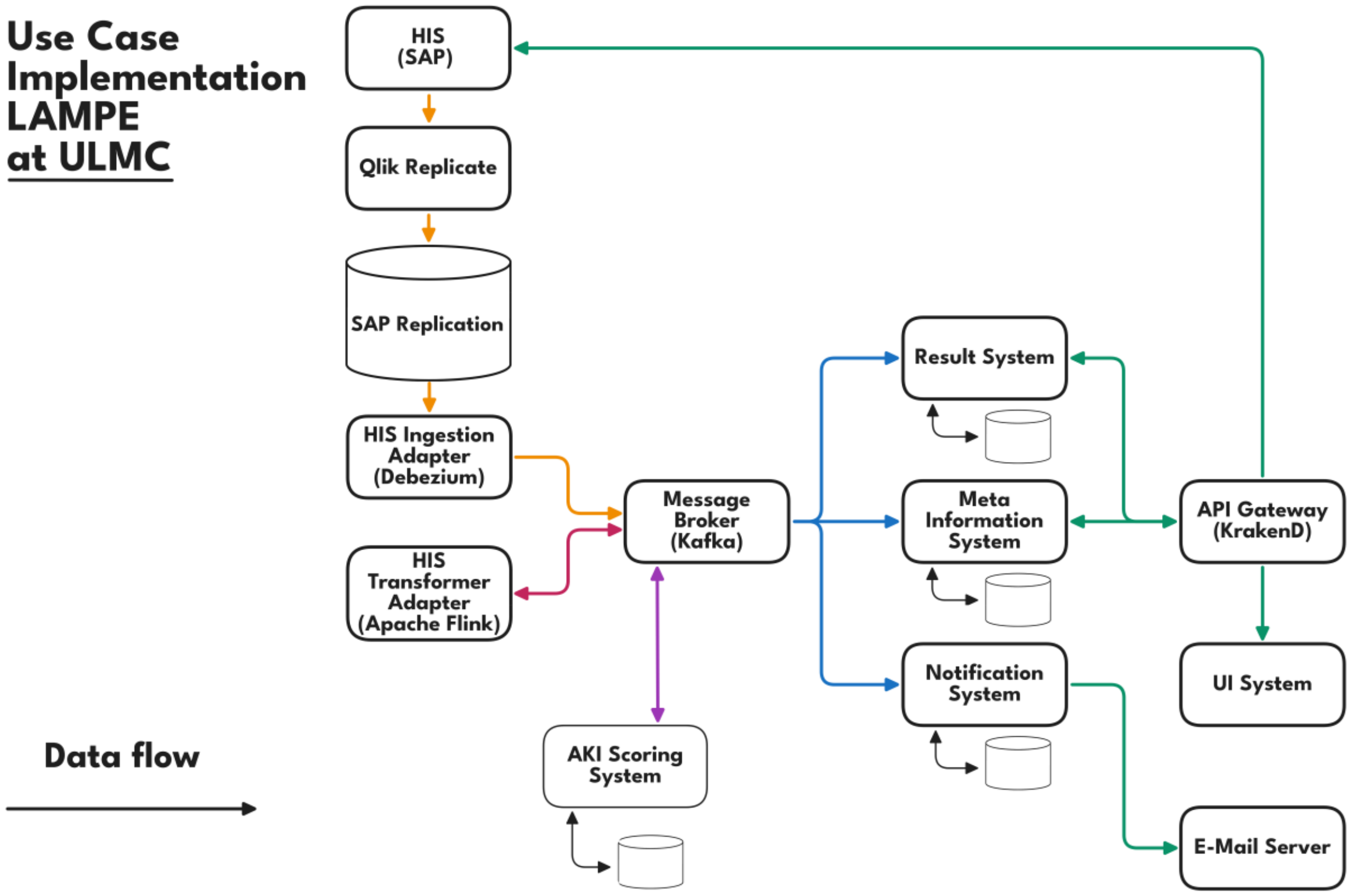

***Figure 5.*** *Overview of LAMPE's architecture. Inheriting the platform's basic architecture and extending it with a dedicated data ingestion and transformation pipeline to fit the targeted environment. Components related to monitoring were left out for visual clarity.*

### Linking SAP to the Platform

In order to provide a steady influx of routine care data into the CDSS we had to integrate *SAP i.s.h.med*. However, due to in-house policies it was not possible to leverage SAP's ***MCI*** *(Message Communication Infrastructure)* to push care data directly to downstream systems. The same policies also prohibited custom non-SAP certified software to directly request data. Thus, we chose to employ *Qlik Replicate* (QlikTech International AB), an SAP certified commercial tool for powering data replication at large scale. It is used to continuously mirror SAP's data into a separate MSSQL database that we have access to. This database was configured with ***CDC** (Change Data Capture)* being enabled. Doing so allowed us to immediately react to database changes (inserts, updates etc.) by utilizing *Debezium*[14], a distributed platform for change data capture. *Debezium* was configured to monitor relevant database tables and write every observed change to a dedicated Kafka topic. For this we re-used the same Kafka cluster that the platform requires as a message broker. This represents the first part of a split data ingestion approach as already outlined in section "HIS agnostic".

Data subsequently had to be transformed into platform specific events as they were introduced in section "Events as Messages". This was done in dedicated jobs carried out by *Apache Flink*[15], a platform for stateful stream processing. Each job was configured to run indefinitely to deal with unbounded data streams. In this context, every resulting event is immediately written to the dedicated *AMPEL* Kafka topic used by the platform. This represents the second part and concludes the entirety of ingesting data into the CDSS.

### Detecting AKI

As of publishing this article, a first scoring system was implemented as part of the CDSS, acting as a reference implementation. It is supposed to monitor care data and detect anomalies indicating a potential ***AKI** (acute kidney injury)*[13], while also taking the administration of vancomycin (a potentially nephrotoxic antibiotic) into account. The scoring system was written in Java, is based on *Spring Boot*[16] and made use of the AMPEL platform's SDK to ease integration into the platform itself.

In order to integrate the scoring system into the platform, it is registered as a distinct consumer for AMPEL events with Kafka's *AMPEL* topic. Consumed events are filtered for relevance for this scoring system, i.e. laboratory events, patient events, medical case events and procedure events. Irrelevant events, on the other hand, are dropped. The platform's transactional inbox implementation is used to de-duplicate incoming messages and guarantee their processing. During processing, each message's medical encounter ID is being used to retain message ordering within each medical encounter.

The AKI algorithm also takes historical information into account. To be specific, the algorithm needs to evaluate care data reaching back as far as seven days. This duration is always relative to the current data point that triggered the evaluation. In case of AKI, every laboratory event carrying information about a measured serum creatinine value is considered such a trigger. However, since the message broker's message stream can only be read continuously, data needs to be stored locally in a dedicated database to allow for historical lookups. For performance reasons, the database size is minimized by deleting older entries as long as events in processed messages allow for it. Examples for such events are the death of a patient, or patient discharge from the hospital.

As described, the scoring system calculates a score every time a trigger event is registered within the stream of messages. Each score is subsequently published to the *AMPEL* topic within Kafka, the same topic that the scoring system is also reading messages from. This is done as part of an event named *AmpelLaboratoryScoreCalculatedEvent* by leveraging *Namastack*[17] as a transactional outbox implementation for Spring Boot. The whole process is embedded within a single transaction to ensure that partial outages do not lead to a loss in messages.

As long as the scoring system is running and registered with the platform's message broker to read platform messages, it is able to calculate and publish scores, i.e. it runs indefinitely.

### Monitoring

The platform does not yet expose any metrics (except Kafka) in order to allow operators to monitor its state. Still, we gathered metrics directly from *Debezium*, *Kafka*, the *replication database* and the *virtual machines* hosting the CDSS. Some of these services already provided metrics via REST endpoints while others required additional software to expose them. Metrics scraped by and stored in a dedicated *Prometheus*[18] instance and visualized in *Grafana*[19]. Using *Alertmanager*, separate rules were employed that are supposed to trigger notifications to the operations team as soon as anomalies were detected within the gathered metrics or partial outages happened. This allowed us to derive the overall flow of information getting ingested from the HIS into the platform and monitor it.

## Discussion

The AMPEL platform includes “medical-device-grade” software, meaning source code and documentation that meet the quality standards for medical device software under the MDR. To the best of our knowledge, it is the first OSS project in Europe to meet this requirement, and the first such project worldwide in clinical decision support.

A number of scientific projects have published CDSS development efforts at the prototype stage[20] or, despite announcing plans to do so, have not released them as open source.[21,22] Open-source hospital information system communities (*OpenMRS*, *OpenEMR*, *GNU Health*, *OSHERA VistA* / *WorldVistA*) have also made repeated attempts to integrate a system-specific CDSS. However, none of these projects has yet been fully implemented.

Three comparable open-source projects remain: *OpenCDS*[23] (University of Utah, USA), *HealthRex CDSS*[24] (Stanford University, USA), and *CQL Evaluation Engine*[25] (HL7 Clinical Decision Support Work Group, USA). All three projects are independent of hospital information systems (HIS) and enable the computation of CDS models. The two HL7-related projects, *OpenCDS* and *CQL Evaluation Engine*, also support relevant interoperability standards (e.g., CQL, CDS Hooks, FHIR). However, these projects focus heavily on the computation of CDS models alone, and a scalable real-time application based on the existing work appears difficult to implement. Furthermore, none of the projects supports use in compliance with FDA or MDR regulations, nor, to our knowledge, is any actively used in routine care. Meeting these requirements is essential for using a CDSS in patient care or in clinical studies evaluating its effects on patients.

The primary strength of the AMPEL platform presented here is its ability to enable healthcare facilities to meet these exact requirements using an in-house CDSS. It creates the conditions for clinical studies that can fill the existing evidence gap. The transparency and reproducibility of the AMPEL platform are also innovative in their own right, as they enable comprehensive independent reviews. Medical professionals can therefore place the utmost confidence in the provided decision support. This trust is not merely desirable; it is a critical factor influencing the success of a CDSS.[26]

### Limitations and Outlook

The main limitation of the platform is the intentional lack of a companion data input module, for example in the form of an ETL pipeline. This module must be programmed separately for each site. We have described the relevant regulatory reasons in an accompanying article about the regulatory concept.[27] At the same time, data integration capabilities vary considerably across healthcare facilities and are highly specific to each site, so custom programming may offer advantages in its own right.

By design, the introduced platform retains all transformed data, ingested from a HIS, for an unlimited amount of time. Since this data is mainly kept in Kafka, which is basically an append-only log, data inherently cannot be erased from it. This presents a particular challenge since patients retain the right to their data according to the General Data Protection Regulation (GDPR), including their data's deletion. Currently, this requires wiping all of the platform's data followed by a full reload from HIS data with the patient’s data left out. Future development demands a different approach that only affects requested data records and strengthens GDPR compliance of the platform as a whole. In this regard, there's already related work[28], also without a focus solely on Kafka.[29] These can serve as a solid starting point in this endeavour.

Another limitation is the governance of the open-source project. An in-house medical device used in routine care requires greater attention to long-term reliability. Consequently, security updates, further development, and community support of the AMPEL platform are solely handled by the Department of Clinical AI and Translational Medicine at ULMC. Since the software is used at ULMC itself, there is an intrinsic interest in effective governance. Given the project's pioneering nature, the governance model should be monitored on an ongoing basis and refined over time if necessary.

The chosen software architecture theoretically enables hundreds of algorithms to run in real time. One limitation of this study is that practical experience with live operation is available for only one algorithm that detects acute kidney injury and a potential vancomycin association at ULMC. As expected, there were no performance issues when utilizing the processing power equivalent to a standard desktop PC. The addition of algorithms from previous and ongoing research projects into the platform is already planned for 2027. As numerous additional algorithms, features, or data connections are added over time, we expect future work to also focus on determining where the performance limits lie.

Other potential improvements are related to how scoring systems are getting linked to the platform. There is a need for giving platform operators more control over these systems to enable advanced features (message routing for A/B testing, rate limiting for messages etc.). Additions like these could be achieved by introducing a *governance layer* to the platform that traffic has to flow through when interacting with any scoring system. It would also act as a central hub for gathering operational metrics from all linked scoring systems in order to gain deeper insight into them and also the platform's status itself.

The AMPEL platform was designed for multi-centre scaling of technical components and scientific evidence. It will only realise its full potential through collaborative efforts. We kindly invite you to join the AMPEL platform community and get involved: https://ampel-platform.eu

## Conclusions

The AMPEL platform is the first open-source project that provides - except for the data input module - all components needed to develop an in-house medical device software. Building on this platform, healthcare facilities can develop and operate their own hospital-wide clinical decision support systems that meet EU regulatory requirements, both for use in patient care and for clinical studies designed to close the existing evidence gap.

## Data and software availability

### Underlying data

The scoring system of the presented use case and its validation dataset are published under the MIT license:  https://github.com/ampel-platform/aki-scoring-system (Release 1.0.0)

### Extended data

Technical documentation of the AMPEL platform and manuals for developing in-house medical device software are published under the MIT license: https://ampel-wiki.eu

### Software availability

The AMPEL platform is published under the MIT license: https://github.com/ampel-platform/ampel (Release 2.0.0)

## AI-use declaration

A generative AI tool (“Translate”, Kagi Inc.) was used to translate the English draft. The translation was human-checked and improved before publication. Two AI search tools (“Assistant”, Kagi Inc.; “Gemini”, Google Ireland Limited) were used as part of the literature review. Programming the user interface was assisted by an AI tool (“Codex” and “ChatGPT”, OpenAI).

No other tasks were carried out with AI tools.

## Competing interests

No competing interests

## Acknowledgements

We would like to thank all former colleagues who initiated and accompanied the AMPEL project.

## Funding information

This project was co-financed by the European Union (NextGenerationEU) and through public funds according to the budget decided by the Saxon State Parliament.